# Self-control of a passively phase-locked Er:fibre frequency comb

Andreas Liehl[1], David Fehrenbacher[1], Philipp Sulzer[1], Stefan Eggert[1], Markus Ludwig[1], Felix Ritzkowsky[1], Alfred Leitenstorfer[1,*], Denis V. Seletskiy[1,2,*]

[1]Department of Physics and Center for Applied Photonics,
University of Konstanz, D-78457 Konstanz, Germany.
[2]Department of Engineering Physics, École Polytechnique de Montréal,
Quebec H3T 1J4, Canada.
e-mail: denis.seletskiy@polymtl.ca, alfred.leitenstorfer@uni-konstanz.de

**Femtosecond frequency combs[1,2] have boosted progress in various fields of precision metrology[3]. Nevertheless, demanding applications such as front-end frequency and time standards[4,5], ultrastable microwave generation[6] or high-resolution spectroscopy[7,8] still necessitate improved stability. The spectral bandwidth and absolute position of individual comb lines are crucial in this context. Typically, both parameters are controlled on short and long time scales by tight locking to external optical and microwave references[9] which represent costly and cumbersome additions to the entire setup. Here, we demonstrate fully self-controlled stabilization of a fibre-based femtosecond frequency comb requiring neither optical nor radio frequency external references. In the first step, this technology allows us to optically eliminate the carrier-envelope phase slip via ultrabroadband difference frequency generation[10,11]. The resulting amplification of intrinsically quantum-limited phase noise from the mode-locked oscillator[12] is elegantly addressed in the second step. We efficiently suppress these excess fluctuations by a direct transfer of the superior short-time noise properties of the fundamental oscillator to the offset-free comb. Our combined scheme provides a high-precision frequency reference operating completely autonomously, thus marking a new era for fibre-based sources in advanced applications[13] ranging from space exploration[14] to tests of the invariability of fundamental constants[15].**

The emission spectrum of mode-locked lasers consists of distinct lines which are separated by the pulse repetition rate $f_{rep}$ and shifted from the origin by the carrier-envelope offset (CEO) frequency $f_{CEO}$. Over the last years, such frequency combs have developed into the most powerful tool of modern optical metrology[4] with many applications requiring maximum control over those two degrees of freedom. Although the first demonstrations of femtosecond frequency combs[1,2] exploited mode-locked Ti:sapphire oscillators, their fibre-based counterparts[16,17] recently gained significant attention[18]. Owing to their robustness and compact setup, these fibre combs show outstanding performance even in harsh environments[9,14,19]. However, the shot noise of relatively low intracavity pulse energy together with propagation of light in a dielectric waveguide activates nonlinear amplitude-to-phase coupling which enhances the timing jitter of the intracavity pulse train[20]. This mechanism increases the phase noise of both the CEO and optical carrier frequencies inside the cavity, hence broadening the linewidth of the optical comb modes. Extracavity cancellation of $f_{CEO}$ by means of difference frequency generation (DFG) enables decoupling of its control from the complex dynamics inside a laser resonator. This technique dramatically increases the system simplicity and long-term stability. Still, it was realized only recently that the spectral width of individual comb teeth is detrimentally affected through nonlinear correlations of the intrinsic optical phase noise emerging during the passive elimination of $f_{CEO}$[12,21]. So far, any subsequent reduction has called for an external reference, thus compromising the inherent simplicity of the setup. In this work, we demonstrate a robust scheme where the optical linewidth of the DFG comb is managed in a fully self-controlled manner by deriving the optical reference from the fundamental master oscillator itself.

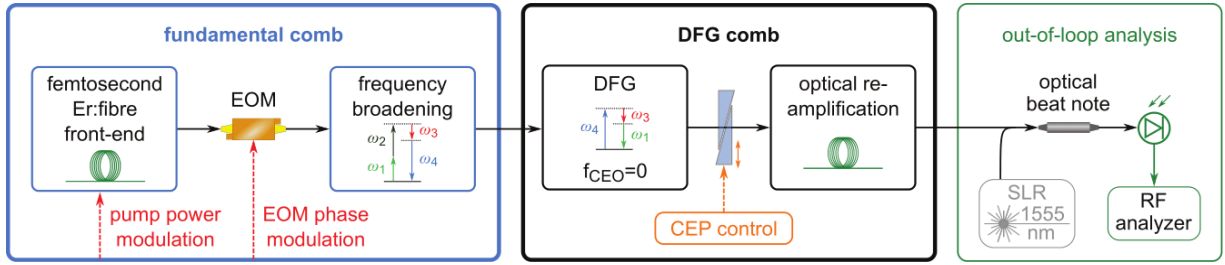

**Figure 1 | Experimental implementation of a passively phase-locked Er:fibre frequency comb.** Left (blue frame): Fundamental comb consisting of a mode-locked Er:fibre front-end and subsequent super-octave frequency broadening. An extracavity electro-optic modulator (EOM) as well as modulation of the pump power of the master oscillator actively suppress optical phase noise. Centre (black frame): Passive elimination of the carrier-envelope offset frequency $f_{CEO}$ via difference frequency generation (DFG). The central output wavelength around 1,550 nm enables optical re-amplification. The carrier-envelope phase (CEP) can be controlled precisely by dynamic insertion of dispersive glass wedges. Right (green frame): Out-of-loop characterization of the optical phase noise and the power spectrum of single comb modes by interference with a low-noise single-line reference (SLR) by means of a radio frequency (RF) analyser.

The basic principle for the generation of a DFG comb at telecommunication wavelengths is sketched in Fig. 1. Extreme frequency broadening of the output of the femtosecond Er:fibre front-end (left blue frame) based on third-order nonlinear processes[22] enables DFG between multiple comb modes which are separated by more than an optical octave. The output centre wavelength around 1,550 nm is matched to the gain bandwidth of Er:fibres for subsequent re-amplification[11] (black frame at centre). Here, no radio frequency (RF) reference is required to lock the CEO frequency to a finite value. The quasi-instantaneous nature of the electronic nonlinearity ensures extraordinary pulse-to-pulse stability of the carrier-envelope phase (CEP)[23]. In contrast, any alternative techniques to eliminate the carrier-envelope phase slip require active control[24,25], thus limiting the effective locking bandwidth.

We study the long-term stability of the CEP by means of a spectrally resolved f-2f interferometer[26]. Fig. 2a depicts the relative frequency stability of $f_{CEO}$ as a function of measurement time $\tau$. Already at $\tau = 1$ s, the relative frequency inaccuracy of the free-running DFG comb is as small as $10^{-17}$ and it decreases below $10^{-19}$ when averaging over approximately 10 minutes. Note that similar performance is provided only by active locking schemes with an extremely optimized feedback loop[27]. We compensate thermal drifts of our laboratory environment accumulated on time scales longer than 20 minutes by a pair of thin glass wedges (see Fig. 1) with actively controlled separation[28]. With this implementation, we readily achieve a relative frequency stability approaching $10^{-22}$ when averaging over a few hours (orange graph in Fig. 2a).

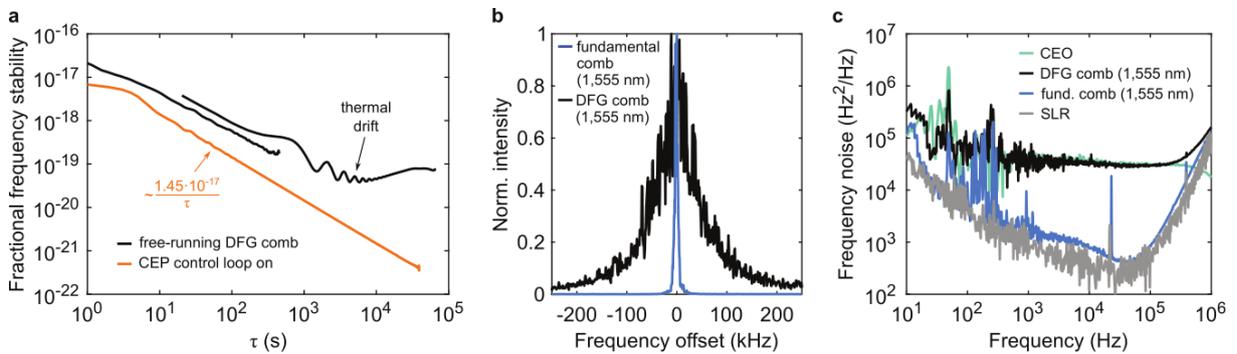

**Figure 2 | Comparison of the performance of the fundamental and DFG comb. a,** Stability of $f_{CEO}$ of the free-running (black) and CEP controlled (orange) DFG comb relative to the detection frequency at 325 THz with respect to the measurement time $\tau$. **b,** Optical power spectrum of comb lines of the fundamental (blue) and DFG (black) comb at a wavelength of 1,555 nm. **c,** Frequency noise spectral density of the fundamental (blue) and DFG (black) comb corresponding to the modes shown in frame b. The frequency noise of the single-line reference (SLR, grey) and a CEO beat note (cyan) are shown for comparison.

These facts motivated us to search for a fully autonomous way for further stabilization. Fortunately, the unique architecture of the passive phase-locking scheme provides access to two frequency combs, namely, the fundamental one and the DFG output (see Fig. 1). For self-controlled stabilization, detailed analysis of the performance of both is required. To this end, we first interfere the fundamental comb with a low-noise single-line reference (SLR) at a wavelength of 1,555 nm. The linewidth of the resulting beat note is 5 kHz (blue in Fig. 2b). This value is dominated by the SLR itself, as indicated by similar frequency noise spectral densities depicted as grey and blue lines in Fig. 2c. Instead, superposition of the DFG comb with the SLR (see right green frame in Fig. 1) reveals a broadening of the optical modes at a wavelength of 1,555 nm to 100 kHz (black graph in Fig. 2b). The noise spectrum is flat and featureless over a large bandwidth (black line in Fig. 2c), confirming the quantum origin[29] of the intrinsic optical phase noise. For comparison, we now determine the CEO frequency noise by interfering the outputs of the DFG process and the fundamental oscillator (cyan line in Fig. 2c). Interestingly, the result of this measurement is almost identical to the optical frequency noise density of the DFG comb (black graph in Fig. 2c). This finding indicates that the CEO beat note contains all information on the intracavity Gordon-Haus jitter[20] necessary for optical phase noise suppression. In the following, we will exploit this insight for a fully self-controlled scheme to narrow the intrinsic linewidth of the DFG comb.

The implementation of our idea is sketched in Fig. 3. We apply two actuators for optical phase noise reduction, both showing negligible cross-talk[12]: an extracavity EOM and modulation of the pump power of the master oscillator (see Fig. 1). The latter is necessary since the EOM does not provide sufficient dynamic range of phase shift for complete elimination of the amplified timing jitter. Nonetheless, implementation of the EOM outside the laser resonator perfectly matches the requirement to address only the timing jitter but not the finite CEO frequency of the mode-locked oscillator. In contrast, modulation of the pump power affects both. Therefore, absolute locking of the CEO beat note via both actuators is not appropriate to purely suppress the enhanced Gordon-Haus jitter efficiently. Thus, we apply two phase-locked loops which generate a proper feedback signal from the CEO beat note: The first one samples $f_{CEO}$ and dynamically changes the input of the second one which accordingly only corrects for phase deviations from the current mean. This technique finally allows dramatic reduction of the optical linewidth as demonstrated by the black and red graph in Fig. 4a, representing an optical comb mode of the free-running and self-controlled DFG comb, respectively. The full width at half maximum (FWHM) of the latter is as low as 5 kHz. Note that the spectral shape of the comb lines is determined by the interference with our single-line reference (see Fig. 1). Hence, this measurement is independent of our self-controlled feedback loop ('out-of-loop'). In the present case, it is clearly limited by our measurement reference whose power spectrum shows a similar FWHM (blue line in Fig. 4a). This conclusion is further strengthened by the analysis of the frequency noise spectral densities, as depicted in Fig. 4b. According Ref. 30, only those components (red shaded regions) which are above the so-called β-separation line (grey dotted) can contribute to the FWHM of the optical comb modes. The frequency noise of the self-controlled comb (red) and the optical reference (blue) show a match below 5 kHz, thus governing the optical linewidth. Contrastingly, features at 50 kHz and 200 kHz induced by the feedback loop manifest themselves as sidebands (red arrows in Fig. 4a) which are suppressed by more than 8 dB with respect to the main peak.

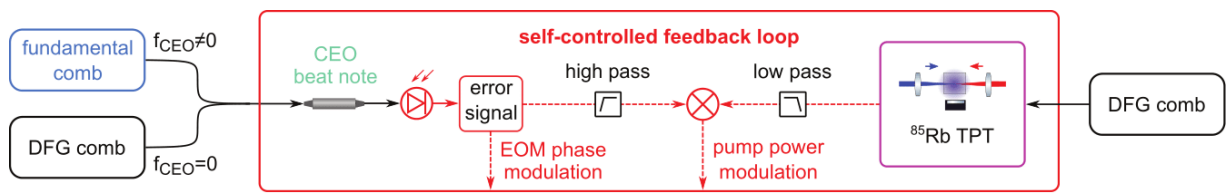

**Figure 3 | Principle of the self-controlled feedback loop.** Superposition of the fundamental and DFG combs results in a CEO beat note (left) which provides feedback to reduce the optical phase noise. The error signal is split to seed both the EOM and the high-frequency modulation port of the pump power of the Er:fibre master oscillator (centre). Slow drifts of the repetition rate are addressed by an additional feedback loop based on the excitation of a two-photon transition (TPT) in $^{85}$Rb (right) controlling the pump power as well.

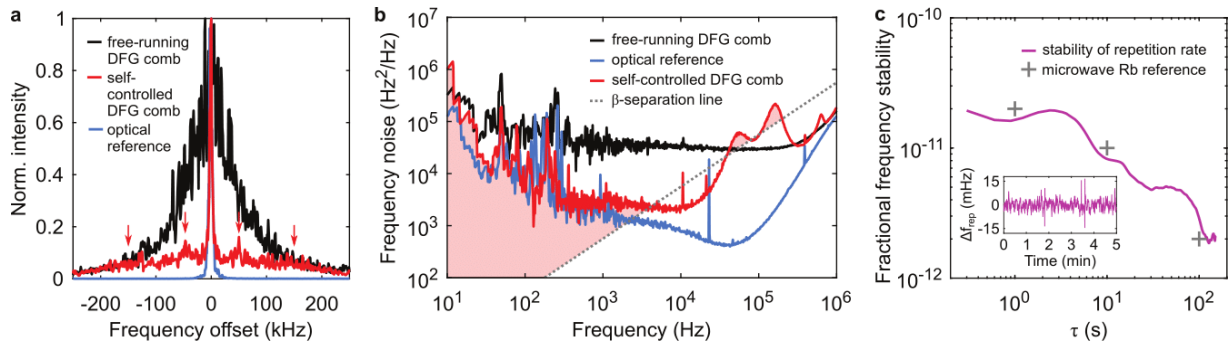

**Figure 4 | Performance of the self-controlled DFG comb. a**, Spectral line shape of comb modes of the free-running (black) and self-controlled (red) DFG comb at a wavelength of 1,555 nm. The power spectrum of our optical reference used for out-of-loop detection is shown in blue for comparison. **b**, Frequency noise spectral density of the free-running (black) and self-controlled (red) DFG comb. To estimate the linewidth, the optical reference (blue) and the β-separation line (grey dotted) are added. **c**, Fractional frequency stability of the pulse repetition rate (purple) locked to an optical two-photon transition in $^{85}$Rb, as measured with a fundamental gate time of 300 ms. The grey crosses represent the manufacturer specification of our microwave measurement reference. Inset: Time evolution of deviations of the locked repetition rate during a time interval of 5 minutes.

In order to obtain a well-defined frequency ruler with an absolute precision, simultaneous compensation for slow drifts of the repetition rate remains as a final task. This goal may be accomplished by Doppler-free excitation of an optical two-photon transition (TPT) in $^{85}$Rb and detection of an appropriate fluorescence signal[12]. Since the Gordon-Haus jitter and slow thermal drifts of $f_{rep}$ are of independent origins and occur on entirely different timescales, the TPT error signal can seed the pump power of the master oscillator if both feedback signals are properly combined within a home-built electronic mixer (see Fig. 3 and Methods for further details). The fractional frequency stability of the repetition rate directly locked to $^{85}$Rb is represented by the purple graph in Fig. 4c, as detected with a dead-time-free frequency counter referenced to a commercial microwave Rb atomic clock. The resulting inaccuracy is clearly limited by the microwave reference (grey crosses, manufacturer specification). The inset shows the temporal evolution of deviations of the repetition rate $\Delta f_{rep}$ from its mean during a time frame of 5 minutes. The optical linewidth of the self-controlled comb remains narrow during the entire measurement interval. Any further long-term stabilization would only require removal of the thermal drift.

In conclusion, we have demonstrated full stabilization of femtosecond DFG frequency combs without employing any external reference. Our self-controlled scheme combines all-passive elimination of $f_{CEO}$ with the synthesis of sharp comb teeth of accurately uniform spacing. The first part restricts the relative frequency inaccuracy of the vanishing CEO frequency of the DFG comb to the $10^{-22}$ regime. Harnessing the fundamental mode-locked oscillator as an ultralow-noise optical reference in the second part alleviates the need for costly ultralow expansion cavities. We emphasize broad applicability of our technique to any frequency comb implementation, regardless of the active lasing medium, the design and the quantum noise level of the master oscillator. Hence, these sources hold great promise as tools of next-generation high precision metrology with a wide accessibility in both commercial and fundamental scientific applications.

## Methods

**Er:fibre laser system.** Our femtosecond Er:fibre front-end (see Fig. 1) consists of an Er:fibre master oscillator with a repetition rate of 100 MHz. Passive mode-locking is achieved by a semiconductor saturable absorber mirror which is contacted to a fibre pigtail in order to avoid any free-space sections. The output spectrum is centred at 1,556 nm and shows a clear solitonic shape. After the oscillator, the pulse energy is boosted above 6 nJ with a single-pass single-mode pumped Er:fibre amplifier. The output pulse is compressed to a pulse duration of 115 fs. Subsequently, an ultrabroad optical spectrum with two intensity maxima separated by 193 THz is generated in a dispersion-managed highly nonlinear fibre[22]. This step enables DFG at an output wavelength of 1,550 nm within a lithium niobate crystal of a periodical poling period adjustable between 21 μm and 35 μm. The resulting phase-stable pulses are re-amplified beyond 6 nJ in a second Er:fibre amplifier and recompressed to a duration of 120 fs. Details of the generation of the DFG comb have been described previously[12]. For comparison, we have built a second laser system with similar pulse properties operating at a repetition rate of 80 MHz[28]. The measurement with additional control of the CEP based on dispersive glass wedges has been performed with this 80 MHz DFG comb.

**Beat note and optical frequency noise measurement.** The beat notes between our single-line reference and the frequency combs (Figs. 2b and 4a) are measured by directing the output of the interferometer to a standard fibre-coupled InGaAs photo diode (Fig. 1, right green frame) and monitoring the RF spectral power on an RF spectrum analyser. The resolution bandwidth is set to 1 kHz and the sweep time is 18 ms and 12 ms for the beat notes of the SLR with the DFG and fundamental comb, respectively. In each case, we have adequately attenuated the power of the SLR in order to avoid saturation of the photo diode.

The frequency noise spectral densities (Figs. 2c and 4b) are measured by observing a 400 ms time frame of the same interferometer outputs on a fast digital oscilloscope with proper sampling. We calculate the noise spectral density with a modified Takeda algorithm[31]. To this end, we first Fourier transform the time evolution of the beat note and select positive frequency components. Subsequently, the time-domain phase deviations are calculated by extracting the phase of the inverse Fourier transform while correcting for phase jumps and subtracting the linear contribution from the carrier. The noise spectral density is finally calculated based on the time-domain phase deviations by Welch's method. We have averaged over ten independent measurements. The frequency noise data of the SLR is provided by the manufacturer.

**Doppler-free excitation of the $5S_{1/2} - 5D_{1/2}$ two-photon transition in $^{85}Rb$.** In order to lock the pulse repetition rate directly to an atomic frequency standard, we exploit the $5S_{1/2} - 5D_{1/2}$ transition in $^{85}Rb$ which is promising for Er:fibre-based lasers since their energy separation is four times the fundamental photon energy at a wavelength of 1,550 nm. It is dipole-forbidden but can be excited in a two-photon absorption process. Therefore, we frequency-double the output of the DFG comb in a periodically poled lithium niobate crystal. The spectrum of the second harmonic contains up to 200 mW of average power. It is split equally around the centre wavelength of 778 nm. The central part of the second harmonic spectrum is blocked and the low and high frequency components are tightly focused into an $^{85}Rb$ gas cell from opposite directions in order to avoid Doppler broadening of the intrinsic linewidth of the TPT. After the excited $^{85}Rb$ atoms have relaxed to the $6P_{3/2}$ level non-radiatively, the strength of two-photon absorption is monitored by detecting the visible fluorescence emitted by the decay to the $5S_{1/2}$ ground state with a photo multiplier tube[12].

**Home-built electronic mixer for feedback to the pump diode laser.** The home-built electronic mixer combines the feedback signals from the CEO beat note to reduce the optical phase noise and the TPT in $^{85}Rb$ to compensate for thermally-induced long-term drifts of $f_{rep}$.

It consists of two input channels which first separately pass impedance converters and a passive low and high pass filter of second order, respectively. Varying the resistance of their RC circuit allows for fine-tuning of the cut-off frequencies. In our experiment, the response function of both the high and low pass filter are adjusted to show a 3-dB roll-off at 50 Hz. After further impedance conversion, both inputs are combined within an additive operational amplifier which provides the final feedback signal for pump power modulation of the diode laser seeding the mode-locked oscillator.

**Acknowledgements**

The authors acknowledge support from the European Research Council (Advanced Grant no. 290876).